\documentclass[10pt,conference]{IEEEtran}
\makeatletter

\let\proof\@undefined
\let\endproof\@undefined

\let\labelindent\@undefined
\makeatother
\usepackage[export]{adjustbox}
\usepackage{epstopdf}
\usepackage[numbers,compress]{natbib}
\usepackage{color}
\usepackage	{setspace} 
\usepackage{array}
\usepackage{lipsum}
\usepackage	{times}
\usepackage	{verbatim}
\usepackage	{graphicx}
\usepackage	{amstext}
\usepackage	{amsmath}
\usepackage	{amsthm}
\usepackage	{amssymb}
\usepackage{wrapfig}
\usepackage[utf8]{inputenc}
\usepackage[vlined,ruled,linesnumbered,commentsnumbered]{algorithm2e}
\usepackage[noend]{algpseudocode}
\usepackage{wrapfig}
\usepackage	[linesnumbered,ruled,vlined]{algorithm2e}
\usepackage	{array}
\usepackage	{footnote}
\usepackage	{url}
\usepackage	{colortbl}
\usepackage	{subfigure}
\usepackage	{multirow}
\usepackage	{caption}
\usepackage {pifont}
\usepackage	{enumitem}
\usepackage{lipsum}
\usepackage[left=1.7cm,right=1.7cm,top=1.7cm,bottom=1.7cm]{geometry}

\title {\huge A Variable Length Coding Framework for Cost Function Reduction in Non-Volatile Memory Systems}
\author{\IEEEauthorblockN{Seyed Mohammad Seyedzadeh,~Alex K. Jones,~Rami Melhem}
\IEEEauthorblockA{\small{seyedzadeh@cs.pitt.edu,\{akjones.melhem\}@pitt.edu\vspace{-1.5em}}
\vspace{-1pt}{}\\
\vspace{0pt}}
}

\begin{document}
\pagestyle{plain} 

\maketitle
\makeatletter
\pagenumbering{arabic}
\begin{abstract}
Variable length coding for Non-Volatile Memory (NVM) technologies is a promising method to improve memory capacity and system performance through compressing memory blocks. However, compression techniques used to improve capacity or bandwidth utilization do not take into consideration the asymmetric costs of writing 1's and 0's in NVMs. Taking into account this asymmetry, we propose a variable length encoding framework that reduces the cost of writing data into NVM. Our experimental results on 12 workloads of the SPEC CPU2006 benchmark suite show that, when the cost asymmetry is 1:2, the proposed framework is capable of reducing the NVM programming cost by up to 24\% more than leading compression approaches and by 12.5\% more than the flip-and-write approach which selects between the data and its complement based on the programming cost.\footnote{This work is supported by NSF grants CCF-1064976 and an SGMI grant from Samsung electronics.}\vspace{-0.5em}
\end{abstract}
\section{Introduction and Motivation}
Non-Volatile Memory~(NVM) technologies such as Spin-Transfer Torque Random Access Memory (STT-RAM) and Phase Change Memory (PCM) are emerging as promising replacements to traditional SRAM cache and DRAM in different levels of the memory hierarchy. STT-RAM is normally employed as a competitive replacement of SRAM as on-chip memories because of its fast read speed, low leakage power, and high density~\cite{nigam2011delivering,seyedzadeh2015pres,seyedzadeh2016improving}. PCM is widely studied as a potential candidate of the main memory due to its higher density and lower standby power compared to DRAM~\cite{smullen2011relaxing}.

A common feature across these NVM technologies is the substantial read-write asymmetry, with write operations taking longer time and more energy than read operations and often more time and energy than the equivalent write operations in other technologies~\cite{zhang2012asymmetry}. For example, the reset energy in STT-RAM is twice as expensive as the set energy. In addition to the asymmetry in energy and programming time, writing a bit value of 0 on PCM cells decreases the cell endurance more than writing a bit value of 1~\cite{zhang2009characterizing}. This feature of asymmetric cost may be leveraged by coding approaches to reduce the programming cost, with the cost being a metric related to energy efficiency, endurance and bandwidth.

Several data compression approaches, including the use of variable length coding~(VLC), have been proposed to improve capacity, bandwidth utilization or programming cost. For example, compression using Frequent Pattern Compression (FPC) or Base-Delta-Immediate Compression (BDI)~\cite{pekhimenko2012base}, traditionally used to achieve efficient capacity and decrease off-chip bandwidth, have been proposed for saving energy. Specifically, compression can be used to reduce the number of written bits while padding the rest of the data block with ``don't cares'' that do not have to be written. Keeping the size of the data block fixed in memory avoids the complexity of managing and locating variable length data blocks~\cite{li2015space}.  Unfortunately, all compression techniques are oblivious to asymmetric costs and treat bit values of zeros and ones equally. 

Compared to compression techniques, fixed length coding approaches may take the writing cost into account via extending the code space. For example, Flip-N-Write (FNW)~\cite{cho2009flip} is a simple encoding technique that may save energy by decreasing the number of reset operations at the expense of increasing, rather than decreasing, the total number of written bits.  A technique is needed that both reduces bit written and uses encoding to improve asymmetric writing costs.

%
\begin{figure}[!b]
\begin{center}
\includegraphics[width=\linewidth]{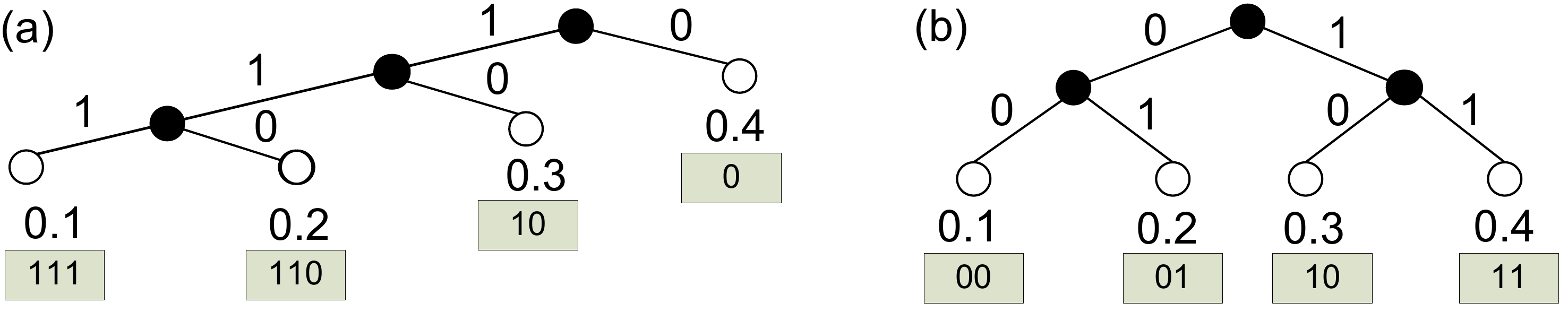}
\caption{\small (a) Huffman encoding tree, (b) A symmetric encoding tree. The floating numbers represent the frequencies of 00, 01, 10 and 11 in the data.
}\label{fig1}
\end{center}\vspace{-1em}
\end{figure}
%
\section{Contributions}
The core idea of this paper is the use of VLC-based compression and encoding to address both bits written and bit-encoding to minimize cost.  We employ a tree-based code whose structure is correlated to the hardware complexity and cost, where the cost of writing a memory block
is $C=n_{0}\times \alpha_0 + n_{1}\times \alpha_1$ such that $n_{0}$ and $n_{1}$ are the number of written zeros and ones, respectively, and $\alpha_0$ and $\alpha_1$ are the costs writing zeroes and ones, respectively. Note that with compression, $n_0 + n_1$ is smaller than the number of bits in the original memory block (\emph{i.e.}, it is compressed).

To make the idea more concrete, consider the encoding trees depicted in  Figure~\ref{fig1} and assume that the relative frequencies of having `00', `01', `10' and `11' in the memory block are 0.1, 0.2, 0.3 and 0.4, respectively. Figure~\ref{fig1}(a) shows the Huffman binary tree which maximizes compression with the code words at the leaves selected to minimize the cost assuming that $\alpha_0 > \alpha_1$. The average code length in this Huffman code is 1.9, with 0.9 zeroes and 1.0 ones, on average. Figure~\ref{fig1}(b) shows a balanced encoding tree with the encoding at the leaves also selected to minimize the cost assuming that $\alpha_0 > \alpha_1$. The average code length in this fixed length code is 2.0, with 0.7 zeroes and 1.3 ones, on average. 

If each two consecutive bits in a memory block is encoded using the above codes, the cost of using the encoding of Figure~\ref{fig1}(a) will be proportional to $0.9\times \alpha_0 + \alpha_1$ while the cost of using the encoding of Figure~\ref{fig1}(b) will be proportional to $0.7\times \alpha_0 + 1.3 \times \alpha_1$. With simple arithmetic, we can find that the cost of the Huffman encoding is lower than the balanced encoding when $\alpha_0 < 1.5 \alpha_1$, and is higher otherwise.  For example, when $\alpha_0 = 2 $ and $\alpha_1 = 1$, the cost for the Huffman encoding is 2.8 and for the balanced encoding is 2.7. On the other hand, when $\alpha_0 = \alpha_1 = 1$, the cost for the Huffman encoding is 1.9 and for the balanced encoding is 2. 

\begin{figure}[!t]
\begin{center}
\includegraphics[width=0.8\linewidth]{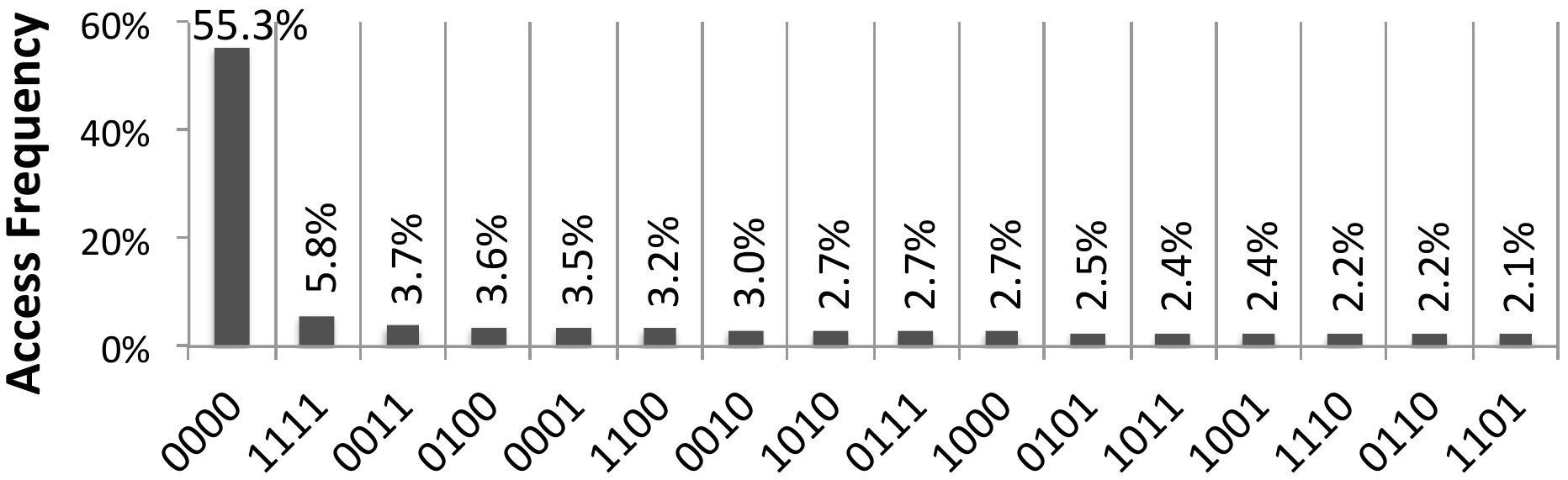}
\caption{\small The percentage of access frequency of 4-bit data words in the SPEC2006 suite benchmark.}\label{fig2}
\end{center}\vspace{-1.75em}
\end{figure}
Given the structure of the encoding tree, the frequency of each encoded symbol and the costs $\alpha_0$ and $\alpha_1$, it is straight forward to assign code words to symbols in a way that minimizes the average code cost (as was done in Figure \ref{fig1}). In the case of 4 symbols (code words), the encoding tree has 4 leaves, and there can only be two structurally different binary trees with 4 leaves\footnote{Interchanging the right and left subtrees of any node in a tree results in a structurally equivalent tree.}. Unfortunately, the number of structurally different trees grows exponentially when the number of leaves increases. For example, there are 10905 structurally different binary trees with 16 leaves~\cite{francisco2015catalan,online1}. In order to select a tree which minimizes the cost of the code, we need to generate various binary tree candidates and select the one leading to the least cost. However, a trade-off should be made between the benefits of reducing the cost of a code and the hardware complexity of the encoder/decoder for that code.
\begin{table}[!b]
\centering\vspace{-0.5em}
\caption{\small Mapping between data words and code words.\vspace{-0.5em}}
\setlength{\tabcolsep}{0.4em}
\footnotesize
\label{tab:1}
\begin{tabular}{|l|l|l|l|l|l|l|l|l|}
\hline
dataword & 0000 & 0001 & 0010 & 0011 & 0100 & 0101 & 0110  & 0111 \\ \hline
codeword & 111  & 0101 & 1100 & 1101 & 1011 & 0100 & 00001 & 0110     \\ \hline
dataword & 1000 & 1001 & 1010 & 1011 & 1100 & 1101 & 1110 & 1111 \\ \hline
codeword & 0011 & 0010 & 1001 & 0001 & 1010 & 00000 & 1000 & 0111      \\ \hline
\end{tabular}
\end{table}

To illustrate our proposal for using VLC for reducing the cost of writing in asymmetric memories, we used 12 workloads from the SPEC2006 suit benchmark to analyze the access frequency of 4-bit patterns in the workload's data. Our analysis in Figure~\ref{fig2} shows that `0000', on average, constitutes about 55\% of the 4-bit patterns in the workloads. In contrast, the percentage of frequencies of other 4-bit patterns range from 2.1\% to 5.8\%. 
Since, on average, more than 50\% of data inputs include `0000', its mapping to a code word shorter than 4 frequently compresses the data block. Clearly, the selection of a code word with very small length such as 1 or 2 for `0000' reduces the number of written bits. However, this increases the diversity in the lengths of the code words, thus, increasing the hardware complexity of the encoding and decoding processes. 

Given the information in Figure~\ref{fig2}, and taking into consideration the encoder/decoder complexity, we show in Table \ref{tab:1} a VLC for encoding every 4 bits in a memory block. The code was determined using a heuristic that builds structurally different 16-leaf code trees with heights between 3 and 5 to restrict the length of the code words to be between 3 and 5 bits. For each tree, the code words are selected in order to minimize the average cost of the code words given the frequencies  from Figure~\ref{fig2} and assuming that $\alpha_0 = 2 \alpha_1$. As a result, the data words with higher frequencies and higher writing cost are mapped to low-cost code words with smaller~(or equal) sizes and vice versa. Finally, the encoding tree that results in the lower average cost is selected.

The selection of the tree that reduces the cost function is chosen off-line and is stored in a look-up table. In the encoding process, the input data block is partitioned into 4-bit groups and each group is mapped to the corresponding code word according to Table~\ref{tab:1}. The mapping of the data words to the corresponding code words can be done serially or in parallel, resulting in a trade-off between latency and area overhead. As long as the length of the produced encoded data is less than the original data block size, the encoded data is written into the memory (as an integer number of bytes); otherwise, the original data block is written. To retrieve the original data during decoding, a dirty bit is used in the encoding process to indicate whether the stored data is encoded or not.

\section{Evaluation and Results}
\begin{figure}[!t]
\begin{center}
\includegraphics[width=0.9\linewidth]{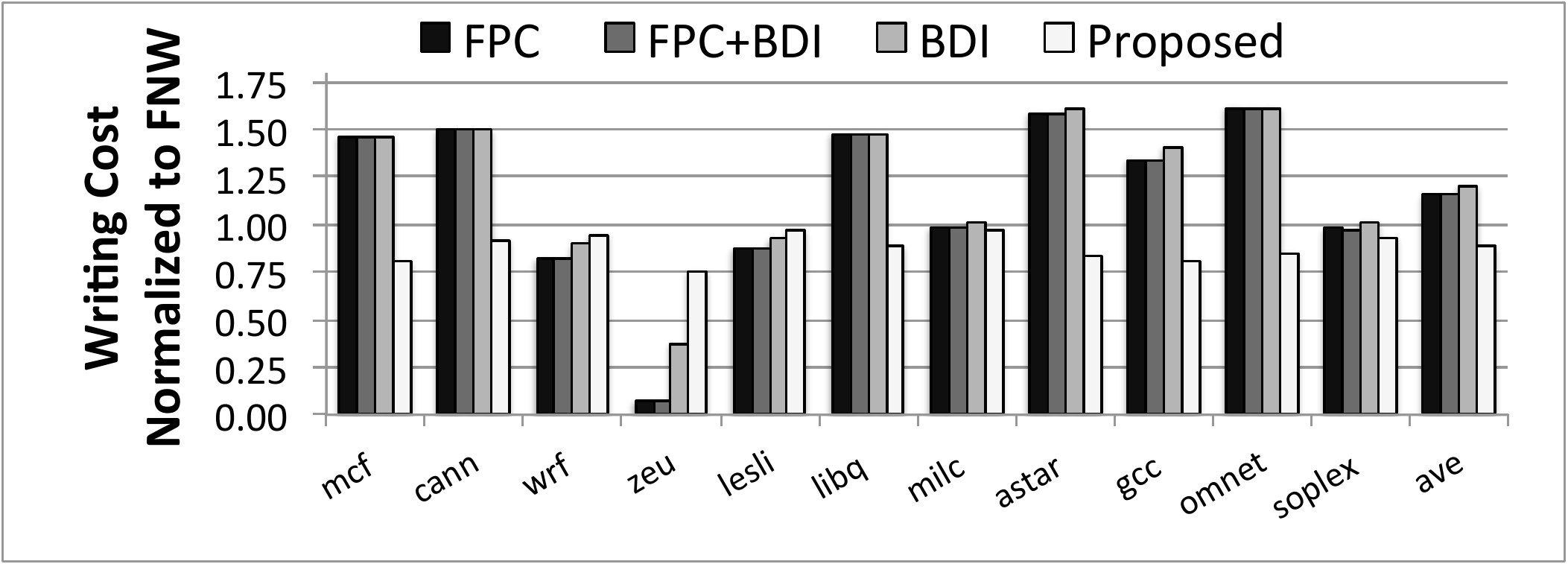}
\caption{\small Writing cost normalized to FNW. The writing cost of bit values of 0 and 1 are 2 and 1, respectively\cite{lee2010phase}.}\label{fig4}
\end{center}\vspace{-1.75em}
\end{figure}

We evaluated the writing cost of the proposed framework against FPC, BDI, FPC combined with BDI and FNW~\cite{cho2009flip} on the 12 workloads from SPEC into an STT-RAM last-level cache. We selected a 512-bit cache-line granularity, and assumed the writing cost of 0 is twice that of a 1 (reasonable for STT-MRAM), in the experimental tests. 
The results, normalized to FNW, are shown in Figure~\ref{fig4}. They indicate that the proposed encoding decreases the writing cost by 24\%, 27\%, 23\%, and 12\% relative to FPC, BDI, FPC+BDI, and FNW, respectively. The cost improvement is mainly a result of mapping costly data words to low-cost code words, which considerably reduces the cost of writing. This demonstrates the potential of our proposal for using VLC to reduce the writing costs in asymmetric NVM, even with the restriction of limited variability in code word lengths.

%
%
\bibliographystyle{IEEEtran}

\end{document}